\documentclass[preprint]{aastex}
\usepackage{amsmath}
\usepackage{graphicx}

\newcommand{\paperone}{Paper~{\sc i}}
\newcommand{\ovi}{O~{\sc vi}}
\newcommand{\rsun}{$R_\odot$}

\slugcomment{accepted version}

\shorttitle{Stereoscopic vs. spectroscopic analysis of a CME}
\shortauthors{Susino, Bemporad \& Dolei}

\begin{document}

\title{3D stereoscopic analysis of a Coronal Mass Ejection and comparison with UV spectroscopic data}

\author{Roberto Susino and Alessandro Bemporad}
\affil{INAF - Osservatorio Astrofisico di Torino, via Osservatorio 20, I-10025 Pino Torinese (TO), Italy}
\email{susino@oato.inaf.it}

\and

\author{Sergio Dolei}
\affil{INAF - Osservatorio Astrofisico di Catania, via S. Sofia 78, I-95123 Catania, Italy}
\email{sdo@oact.inaf.it}

\begin{abstract}
A three-dimensional (3D) reconstruction of the 2007, May 20 partial-halo Coronal Mass Ejection (CME) has been made using STEREO/EUVI and STEREO/COR1 coronagraphic images.
The trajectory and kinematics of the erupting filament have been derived from EUVI image pairs with the ``tie-pointing'' triangulation technique, while the polarization ratio technique has been applied to COR1 data to determine the average position and depth of the CME front along the line of sight.
These 3D geometrical information have been combined for the first time with spectroscopic measurements of the \ovi\ $\lambda\lambda$1031.91, 1037.61~\AA\ line profiles made with the \emph{Ultraviolet Coronagraph Spectrometer} (UVCS) on board SOHO.
Comparison between the prominence trajectory extrapolated at the altitude of UVCS observations and the core transit time measured from UVCS data made possible a firm identification of the CME core observed in white light and UV with the prominence plasma expelled during the CME.
Results on the 3D structure of the CME front have been used to calculate synthetic spectral profiles of the \ovi\ $\lambda\lambda1031.91$~\AA\ line expected along the UVCS slit, in an attempt to reproduce the measured line widths.
Observed line widths can be reproduced within the uncertainties only in the peripheral part of the CME front; at the front center, where the distance of the emitting plasma from the plane of the sky is greater, synthetic widths turn out to be $\sim 25$\% lower than the measured ones.
This provides strong evidence of line broadening due to plasma heating mechanisms in addition to bulk expansion of the emitting volume. 
\end{abstract}


\section{Introduction}

CMEs are large-scale magnetized plasma structures that are ejected from the Sun into the heliosphere over the course of minutes to hours, with a wide range of sizes, masses, and velocities.
CMEs can be associated with the eruption of filaments and/or solar flares or without any apparent trigger \citep[see, e.g.,][]{web12,mor12}. 
The heliospheric counterparts of CMEs are often referred to as interplanetary coronal mass ejections (ICMEs), and they are one of the main drivers of space weather events on Earth \citep{tsu88,zha07}. In fact, ICMEs can produce severe magnetic storms when interacting with the Earth's magnetosphere \citep[see, e.g.,][]{gos93,sch06}.

Many aspects related to the initial phases and subsequent evolution of CMEs are, at present, only poorly understood.
To deepen as much as possible the knowledge of the processes involved in CMEs initiation and propagation is crucial to improve the magnetohydrodynamics (MHD) simulation techniques and, in turn, the space weather forecasting techniques. Significant insights on these problems can be gained by studying the three-dimensional (3D) structure of CMEs. 
For instance, it has been proved that the geo-effectiveness of a CME (i.e., its capability to interact with the Earth's magnetosphere and trigger geomagnetic storms) strongly depends on the magnetic configuration of the associated ICME, in particular on the relative orientation of the magnetic field embedded in the magnetic cloud associated with the ICME with respect to the Earth’s magnetic field \citep[e.g,]{yur01,lan01,bot03}. 
Therefore, a better determination of the 3D structure of CMEs may lead to a better understanding of the possible correlations between the initial evolution of CMEs, the subsequent propagation and magnetic configuration of ICMEs, and their eventual impact on the Earth.

CMEs are typically observed in white-light (WL) images provided by coronagraphs, such as the Large Angle and Spectrometric Coronagraph \citep[LASCO;][]{bru95} on board the Solar and Heliospheric Observatory (SOHO), and the COR coronagraphs on board the Solar Terrestrial Relations Observatory (STEREO) twin spacecraft \citep{kai08}. 
White-light continuum emission arising from the coronal plasma is produced by Thomson scattering of photospheric light by free electrons \citep{min30,bil66}. Since the corona is optically thin, white-light images actually show the projection on the plane of the sky (POS) of the 3D CME structure, integrated along the instrumental lines of sight (LOSs). Moreover, CMEs are not observed in isolation, but in the presence of the fine structural detail of the surrounding corona \citep{mor12}. 
Projection and LOS integration effects can be rather severe and can significantly affect the determination of CME properties \citep[e.g.,][]{bur04}.
For halo and partial-halo CMEs, the projected altitude of CME features inferred from coronagraphic images can notably differ from the actual one; for this reason, real position and speed of generic plasma elements cannot be obtained from such observations unless a geometry of the CME is assumed. However, even in that case limitations and approximations are unavoidable \citep[see, e.g.,][]{cia06}.
The main challenge is therefore to determine the location and the geometry of CMEs in the 3D space using one or more bi-dimensional images, corresponding to different viewpoints of the same CME feature.

The STEREO mission was conceived to properly address these issues; since its launch in 2006, a wealth of methods have been developed and tested in order to determine the full 3D structure of CMEs. 
We refer the interested reader to \citet{mie09,mie10}, who compared the results of different 3D reconstruction techniques applied to STEREO images, \emph{in primis} the triangulation or tie-pointing technique \citep[see, e.g.,][]{inh06} and the polarization ratio technique, originally developed by \citet{mor04}. 
A different approach to estimate the 3D CME structure uses forward modeling, that consists in making strong assumptions on the morphology of the feature that is intended to be reconstructed \citep[e.g., the graduated cylindrical shell flux-rope model of][]{the06,the09}.

Further investigation method of CMEs is based on ultraviolet (UV) spectroscopic observations, such as those provided by the Ultraviolet Coronagraph Spectrometer \citep[UVCS;][]{koh95} on board SOHO. 
These have the potential to give information on both kinematics (e.g., LOS and radial component of the plasma speed from line Doppler shifts and Doppler dimming) and thermodynamics (e.g., kinetic temperature of the emitting ions from line widths and plasma electron density from line intensity ratios) of the CMEs, that are complementary to those coming from stereoscopic analyses. 
In fact, the derivation of some physical plasma parameters from UV spectra, such as the kinetic temperature, is affected by uncertainties related to the unknown 3D distribution of plasma along the LOS. 
These uncertainties can be sensibly reduced, or even eliminated, if the 3D structure of the emitting plasma is determined, for instance, by the stereoscopic images.

Therefore, the study of CMEs through a combined analysis of stereoscopic data and spectroscopic UV observations allows (1) to derive new significant information, such as the contributions to the line broadening due to both the bulk expansion and the heating, whose determination would not be otherwise possible without the combination of the techniques of analyses, and (2) to place important constraints to MHD numerical simulations attempting to investigate the initiation processes and evolution of these events.
Although several works use the above mentioned techniques of 3D reconstruction (i.e., triangulation and polarization ratio) to determine the spatial and geometrical configuration of CMEs and measure their kinematic parameters (e.g., the initial speed or the propagation direction of the CME front with respect to the LOS), no case to our knowledge of the literature presents a direct combination of stereoscopic analysis applied to STEREO images with the results of the 3D geometry of CMEs obtained by UV spectroscopic observations.

In the present work, we used the STEREO/COR1 and SOHO/UVCS images of the partial-halo CME occurred on 2007, May 20. In particular, we applied the polarization ratio technique to the COR1 data in order to determine the 3D structure of the CME front along the LOS. Then, for the first time ever, we exploited the geometric information obtained by the 3D reconstruction to synthesize spectral profiles of the \ovi\ $\lambda1031.91$~\AA\ line and compare the simulated profiles with those measured by UVCS in order to provide constraints on the coronal plasma heating associated with the transit of the CME.
The paper is organized as follows: in \S 2 we illustrate the approach we adopted to select the CME event analyzed here; in \S 3 we describe the coronagraphic (from LASCO and STEREO) and spectroscopic (from UVCS) observations acquired on 2007 May 20; then we summarize results obtained in previous works related to the same event (\S 4.1), we describe how STEREO data have been analyzed to derive the 3D reconstruction of the erupting filament (\S 4.2) and of the CME front (\S 4.3) and how these results have been combined with the spectroscopic observations from UVCS (\S 4.4).
A discussion of our results concludes the paper (\S 5).

\section{Selection of the event}

The main target of the present work is to demonstrate the possible advantages for the reconstruction of a CME event when both stereoscopic and spectroscopic data are available. 
The only instrument that was able to provide daily observations of UV spectra emitted by the intermediate corona is the UVCS spectrometer onboard SOHO \citep[see][]{koh95}, now switched off, while other spectrometers on SOHO (CDS, SUMER) and subsequent solar missions (HINODE/EIS) focused on EUV spectra emitted by the lower corona. 
Data acquired by the latter instruments cannot be used for our purposes, because when the CME falls in the field of view of the spectrometer is out of the field of view of the coronagraphs on board SOHO and STEREO, and vice-versa, making contemporary observations not possible. 
UVCS data are available in the catalog until December 2012, but starting from early 2006 some dark columns began to appear and to broad over the UVCS detectors, thus significantly reducing the region available for scientific analysis on the data and in turn the number of detected CMEs. 
For this reason, the UVCS CME catalog recently released stops at the end of year 2005 \citep[see][]{gio13}. 
On the other hand, the STEREO mission was launched in October 2006, when the UVCS detector already had some problems, thus making very difficult a contemporary observation of the same CME.
For this reason there are no published works dealing with STEREO and SOHO/UVCS data relevant to the same event, as done here.

In order to identify an event observed by UVCS in the STEREO era, we performed an automated cross search of the LASCO CME catalog \citep{gop09} and the UVCS mission log catalog available online from the SOHO archive.
We restricted our research to all the events listed in the CME catalog that have occurred since October 2006 till December 2010.
For each event, we verified if the following conditions were satisfied: (1) UVCS observations were available on the same day; (2) the start (end) of UVCS observations preceded (followed) the time when the CME front reached the altitude of the UVCS field of view; (3) at least 20\% of the UVCS slit overlapped the area spanned by the CME front during its expansion.
Information on the CME time and linear velocity reported in the LASCO CME catalog have been used to calculate the time of the transit of the CME front at the heliocentric distance of the UVCS slit reported in the UVCS mission logs.
The CME central position angle and the CME average angular width have been used to compute the width of the area spanned by the expanding CME front.

The final result is that only one CME event, occurred on 2007 May 20, fulfills these three criteria.
Owing to the more and more degraded state and limited field of view of UVCS detectors since year 2006, it is very difficult that data acquired after December 2010 can be used to study large scale eruptive events like CMEs.
For this reason the event of 2007 May 20 could be the unique possibility we have to perform a detailed combination of stereoscopic and spectroscopic analysis of a CME.
In addition, this event gave origin to an ICME which was observed \emph{in situ} by STEREO and the WIND satellite (see \S 4.1) and which appeared to have interacted strongly with the ambient solar wind, triggering minor geomagnetic storms since May 22, as reported in the May 24 bulletin on solar and geomagnetic activity released by the Solar Influences Data analysis Center (SIDC; {\tt http://sidc.oma.be}). 
Therefore, the analysis of this CME can also provide further insight into Space Weather and its applications.

\section{Observations}

\begin{figure}[p]
\centering\includegraphics[width=\textwidth]{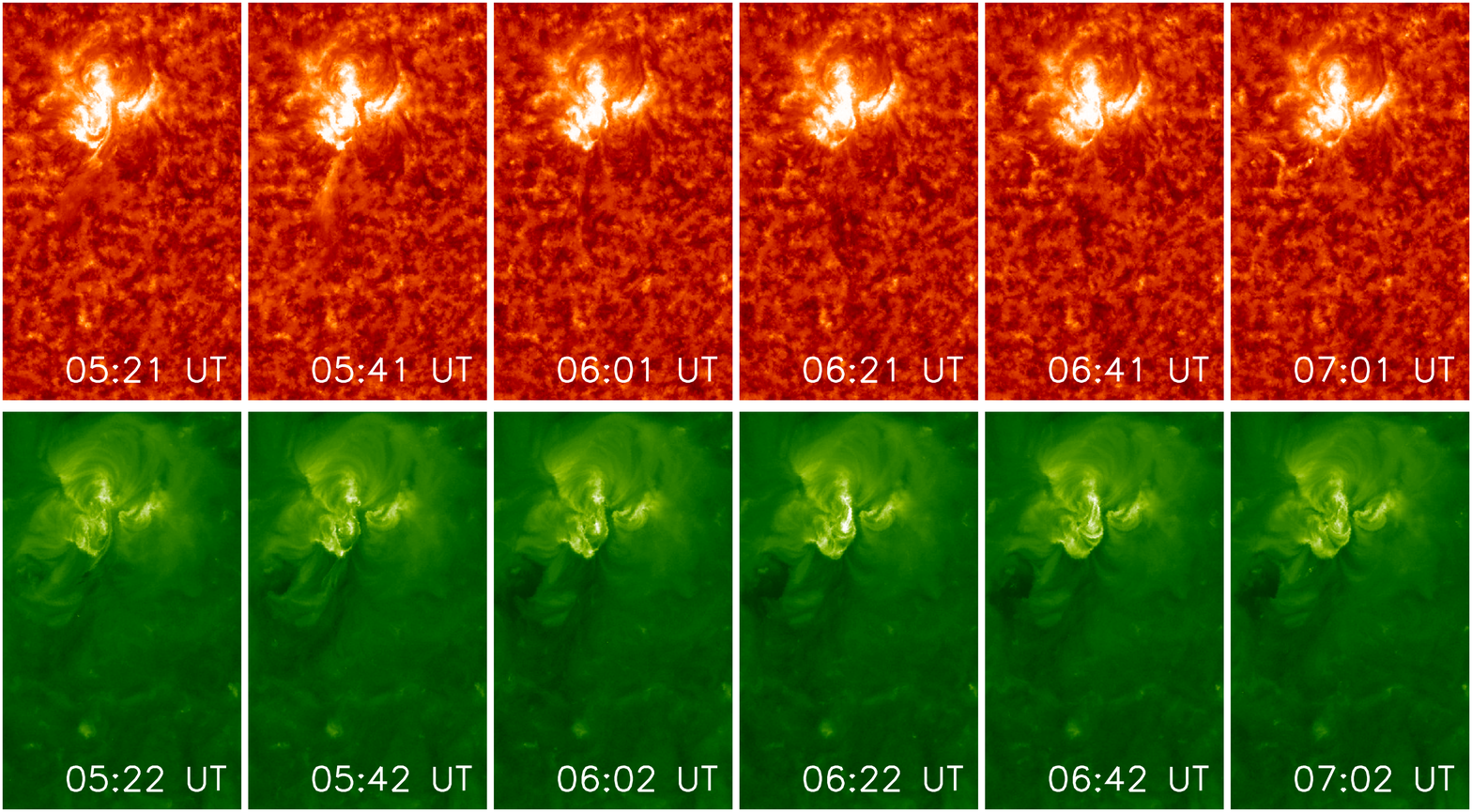}
\caption{Sequence of STEREO/EUVI-B He~{\sc ii} 304~\AA\ (top row) and Fe~{\sc xii} 195~\AA\ (bottom row) close-up images of the source active region of the May 20 2007 CME, at different times during the eruption of the filament given by the labels.\label{euvi}}
\end{figure}

\begin{figure}[p]
\centering
\includegraphics[width=\textwidth]{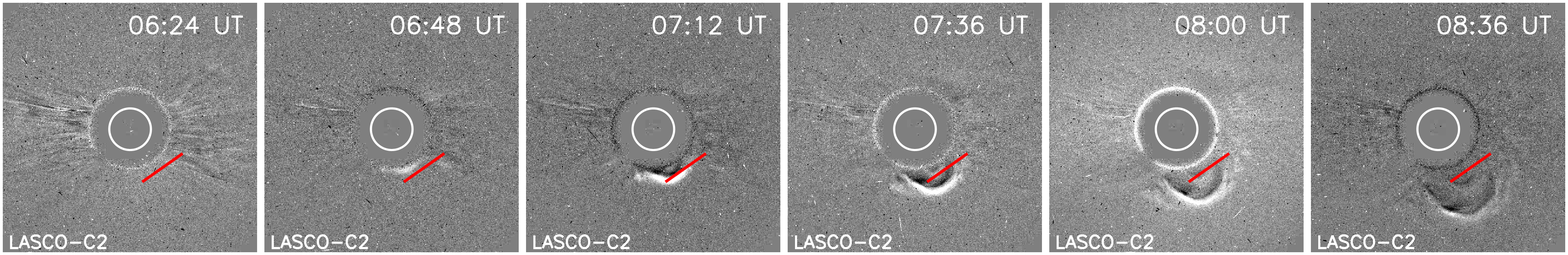}
\includegraphics[width=\textwidth]{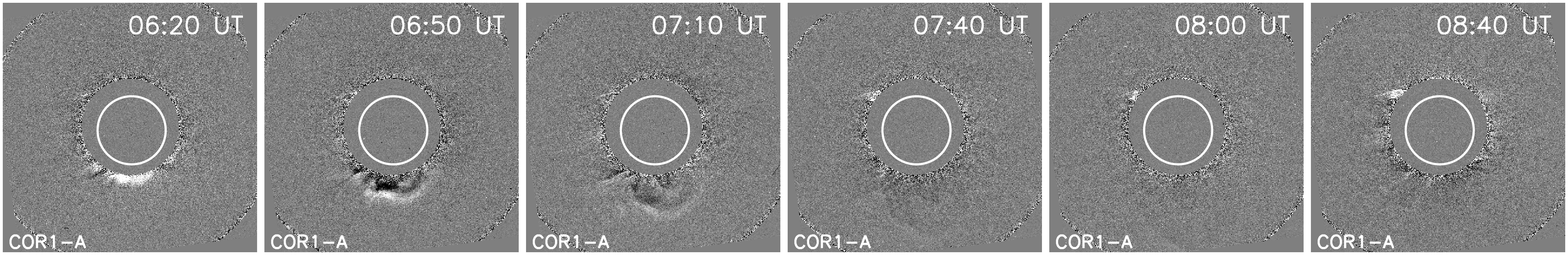}
\caption{Sequence of LASCO-C2 (top row) and COR1-A (bottom row) white-light, running-difference images of the solar corona during the May 20 2007 CME, at the times indicated by the labels. The instantaneous UVCS field of view (the slit, red line) is also superimposed in the LASCO-C2 images for reference purposes.\label{lasco}}
\end{figure}

The 2007 May 20 CME, classified as partial halo in the LASCO CME catalog, occurred after the eruption of a filament from active region NOAA 10596, starting at 04:30~UT.
The event was observed simultaneously by the EUVI and COR1 imagers operating on board the STEREO spacecraft and the UVCS spectrometer on board SOHO. 
At that time the separation angle between STEREO spacecraft was quite small (about 8.7$^\circ$), hence STEREO and SOHO EUV imagers viewed approximately the same portion of the solar disc. 
The source active region (AR) was located close to the disc center and exhibited a quite complex quadrupolar magnetic field configuration, as observed by SOHO/MDI; the eruption was probably triggered by the emergence of a fifth, compact negative flux in the wake of the active region starting about 14~hours before the CME.
The ejection of filament material was observed in the He~{\sc ii} 304~\AA\ line (sensitive to plasma with temperature of $\sim 0.1$~MK) with the Extreme Ultraviolet Imager (EUVI) telescopes on board STEREO, between 04:30~UT and 06:00~UT (see Figure~\ref{euvi}). 
Before the eruption, the filament appeared to be mostly aligned with the North-South direction. During the eruption, an X-ray event of class B9 from the source active region was detected by the GOES instrument, which measured an enhancement of the soft X-ray flux approximately from 05:00 to 08:00~UT, with peak at $\sim 05$:56~UT. 
The flare was visible in the Fe~{\sc xii} line (195~\AA, corresponding to a temperature of about 1.5~MK) with EUVI as a rapid brightening of the AR loop system, starting at 04:52~UT.
The CME leading edge was first detected in white-light at 06:20~UT in STEREO/COR1-A images and at 06:48~UT in SOHO/LASCO-C2 images (see Figure~\ref{lasco}). 
Subsequent images show the characteristic three-part structure observed in many CMEs, with a quite faint, hemispherical front expanding southward at $\sim 275$~km~s$^{-1}$ (as reported in the online LASCO CME catalog), followed by the void and a more dispersed core plasma.

UVCS observations started at 00:38~UT of May 20 and lasted until 17:10~UT.
The slit field of view (FOV) was centered at a heliocentric distance of 2.4~\rsun\ and at a central latitude of 55$^\circ$~S, as shown in Figure~\ref{lasco}.
Spectral profiles were acquired in the wavelength ranges of 1017-1047~\AA, 997-999~\AA, and 953-981~\AA\ with a spatial binning of 8 pixels (56''), spectral binning of 2 pixels (0.198~\AA), and exposure time of 120~s. 
Owing to the poor statistics, only the \ovi\ $\lambda\lambda1031.91$, 1037.61~\AA\ doublet lines had a significant signal-to-noise ratio to be identified.
As stated above, several dark columns are affecting the UVCS detector.
In particular, in our case no signal was recorded in three spatial strips at bins 26, 19-21, and 11-14 (corresponding to latitudes of $57^\circ$S, 48-$51^\circ$S, and 37-$41^\circ$S, respectively).
The total counts relevant to each dark strip are actually accumulated by the detector in one of the adjacent bins (25, 17, and 9, respectively); therefore we were able to recover the signal of each dark column in a strip by replacing that column with the array containing the total counts of the strip divided by the number of bins in that strip.
The result of this procedure, however, is that every column within the same strip will have the same signal (i.e., the same spectrum).
Nevertheless, it is worth noting that just the narrowest of the three dark strips, located at bin 26, falls in the region of the detector crossed by the CME (see below), while the strip at bins 19-21 only marginally affects the outer part of the CME front.
Thus the loss of information caused by this average process is not significant for the purposes of our work.
The spectral profiles of the \ovi\ lines have been integrated after usual standard calibration and background subtraction.
A pre-CME coronal spectrum averaged over the pre-CME exposures has also been subtracted in order to exclude any possible contributions coming from the ambient quiet corona.

\begin{figure}[p]
\centering\includegraphics[width=\textwidth]{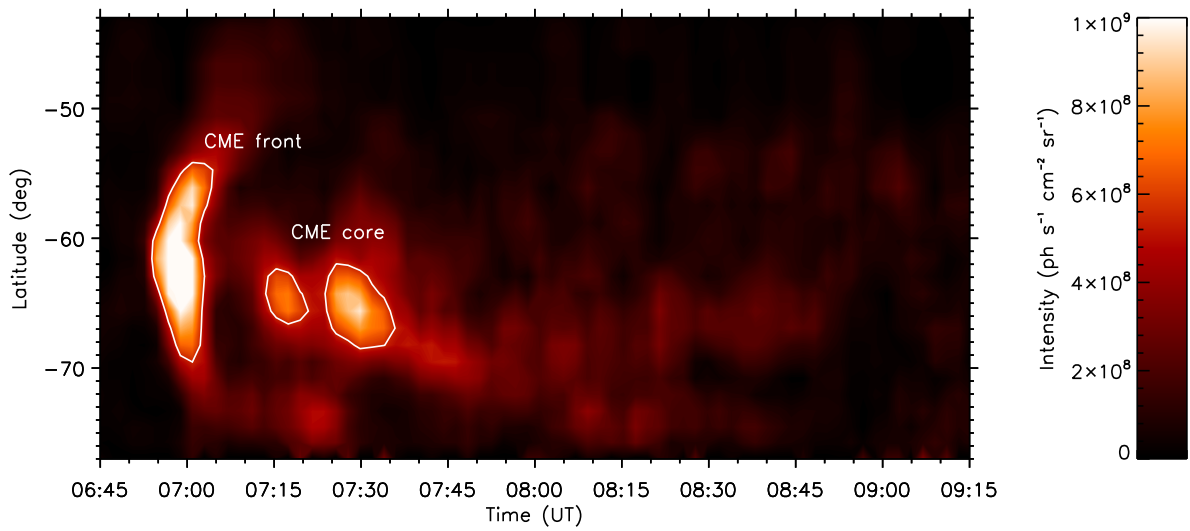}
\caption{\ovi\ $\lambda$1031.91~\AA\ intensity evolution along the UVCS slit ($y$-axis) at different times ($x$-axis) as observed at 2.4~\rsun during the transit of the CME. The typical three-part structure of the CME is clearly visible.\label{uvcs_int}}
\end{figure}

Figure~\ref{uvcs_int} reports a 2D map of the total \ovi~$\lambda1031.91$~\AA\ line intensity observed at different latitudes along the UVCS slit ($y$-axis) and at different times ($x$-axis) over the slit region crossed by the CME.
As it is well known, the emergent intensity in the \ovi\ doublet lines is a combination of collisionally excited and resonantly scattered components.
The collisional component is due to ion excitation by collisions with thermal coronal electrons and depends on the LOS distribution of electron density and temperature.
The radiative component is produced by absorption of chromospheric radiation and depends, in addition, on the plasma flow velocity \citep[see, e.g.,][]{bra13}: when plasma flows are not negligible, the absorption profile of \ovi\ ions is Doppler-shifted with respect to the disk exciting profile and the scattering is less efficient, resulting in a reduction in intensity of the scattered radiation---this effect is known as Doppler dimming \citep[see][]{wit82,noc87}.
If the outflow velocity is sufficiently high to produce significant Doppler-dimming, then the ratio between the intensities of \ovi\ $\lambda\lambda$1031.91, 1037.61~\AA\ line intensities can be used to determine the ion outflow velocity, as originally pointed out by \citet{koh82}.
An approach based on these arguments and widely adopted in the analysis of UVCS data \citep[see, e.g., reviews by][]{ant06,koh06} has been used in our previous paper \citep[][hereafter \paperone]{sus13} to estimate the radial component of the outflow velocity of the plasma embedded in the front and in the core of the CME (see later).
The \ovi\ intensity map reveals the typical three-part structure that is often observed for most of CMEs \citep[see, e.g.,][]{web88,lee06}: the leading loop-like front, which crossed the UVCS slit at 06:54~UT, i.e., about 34 minutes after it appeared in the FOV of COR1, the dark void, corresponding to a sudden drop in the observed \ovi\ emission, and then the core.

\section{Data analysis and results}
\subsection{Summary of previous results}

The 2007 May 20 event has already been studied in earlier works by different authors, with different goals.
\citet{mie08} used a method based on height-time measurements of coronal moving features applied to STEREO/COR1 data to measure the direction of propagation as well as the real speed of this CME.
They estimated an initial unprojected speed of 548~km~s$^{-1}$, whereas the projected velocities measured in STEREO-A and -B spacecraft were 242~and 253~km~s$^{-1}$, respectively.

The interplanetary counterpart of the May 20 CME was studied in detail by \citet{kil09} using \emph{in situ} data from STEREO and the WIND satellite.
The magnetic cloud (MC) associated with this event was observed on May 23 at a distance from the Sun of 0.96~AU (Astronomical Unit, $1~\textrm{AU}\simeq150\times10^6$~km).
The magnetic field configuration of the ICME was derived by solving numerically the Grad-Shafranov equation \citep{hu02} on the STEREO-A data; the orientation of the streamer-belt neutral line and the orientations of the filament axis and the post-eruptive arcades appear to correlate with the orientation of the MC axis.
The estimated leading-edge speed of the MC was 535 km~s$^{-1}$.
An accurate approximation of the ICME transit time made by applying existing models for CMEs propagation through interplanetary space gave the result that if the initial speed of the 20 May CME determined by \citet{mie08} is used, the arrival time of the ICME predicted at the STEREO-A location differs by about 8~hours from the measured one. 

\begin{table}
\begin{center}
	\caption{CME parameters derived from UVCS data (see \paperone)\label{table1}}
	\begin{tabular}{lcc}                       \\
		\tableline\tableline
		& Front & Core     \\
		\tableline
		$v_{\rm POS}$ (km s$^{-1}$) & 289   & 231  \\
		$v_{\rm LOS}$ (km s$^{-1}$) & 500   & 535  \\
		$v$           (km s$^{-1}$) & 578   & 583  \\
		$v_{\rm RAD}$ (km s$^{-1}$) &  75   & 85   \\
		$\theta$      ($^\circ$)    &  30   & 23   \\
		\tableline
	\end{tabular}
\end{center}
\end{table}

In our previous paper (\paperone), we derived from UVCS data the temporal evolution of the \ovi\ line intensities and widths along the direction of propagation of the CME, and measured the velocity components on the POS, along the LOS, and in the radial direction, as well as the \ovi\ kinetic temperature, for both the front and the core of the CME.
These results are summarized in Table~\ref{table1}.
We were also able to estimate the direction of propagation of the CME front with respect to the LOS, using the measured LOS and POS components of the outflow velocity, which turned out to be $\approx 23^\circ$.
These measurements were then compared with first results from the 3D reconstruction of the prominence kinematics using triangulation technique applied to STEREO/EUVI data.
In what follows, we better explain this method and report more complete results on the reconstruction of the erupting prominence.

\subsection{3D reconstruction of the erupting prominence with EUVI data}

The three-part structure of the 2007 May 20 CME clearly visible in the \ovi\ intensity map derived from UVCS data, is typical of many CMEs.
The leading edge, the dark cavity, and the bright core, are generally associated with the compressed solar plasma ahead of the ejecta, the erupting magnetic flux rope, and the cool and dense prominence plasma, respectively \citep[e.g.,][]{cre04,lee06}.
However, a convincing identification of the white-light core of the CME with the cool prominence material is relatively rare and, although recent studies have demonstrated a strong connection between prominence eruptions and CMEs, the correlation is not always one-to-one \citep[see][]{mie11}.
In order to investigate possible correlations between the erupting prominence and the CME structure, we applied one of the available stereoscopic techniques to reconstruct the trajectory and kinematics of the filament and compared the results with those derived from UVCS data and already presented in \paperone.

According to \citet{lie09}, if a CME feature can be identified in both images of a simultaneous pair, such as those recorded by STEREO, then the position of the feature in a three-dimensional heliocentric coordinate system can be determined by triangulation, provided that the location and separation of the two STEREO spacecraft are known.
For a constant error in locating the feature on the 2D images, $\Delta x$, the error in the depth estimate of the 3D reconstruction, $\Delta h$, increases as the separation angle $\phi$ between the two STEREO spacecraft decreases: $\Delta h=\Delta x/\sin(\phi /2)$ \citep[see][]{mie09,lie09}.
However, a reliable identification of the same feature in the two STEREO images can be achieved only if the separation angle is quite small ($\sim 10^\circ$).
For 2007 May 20, the separation angle was $\sim 8.7^\circ$, leading to an error in the height determination of $\approx 1$\%, if the feature could be located to within one pixel. 
The reconstruction results presented in this Section were obtained by manually locating the same feature in two simultaneous STEREO images; this is often referred to as ``tie-pointing'' technique.
As explained in \paperone, the erupting filament was reconstructed applying this technique for 7 different times on 20 May to study its evolution, using running differences of STEREO/EUVI 304~\AA~image pairs in order to enhance the visibility of the prominence. 
Generally, the filament is seen in EUVI 304~\AA, responsive to a temperature of $\sim 0.1$~K, as a dark absorbing feature above the surrounding chromospheric plasma. 

\begin{figure}[p]
\centering\includegraphics[width=\textwidth]{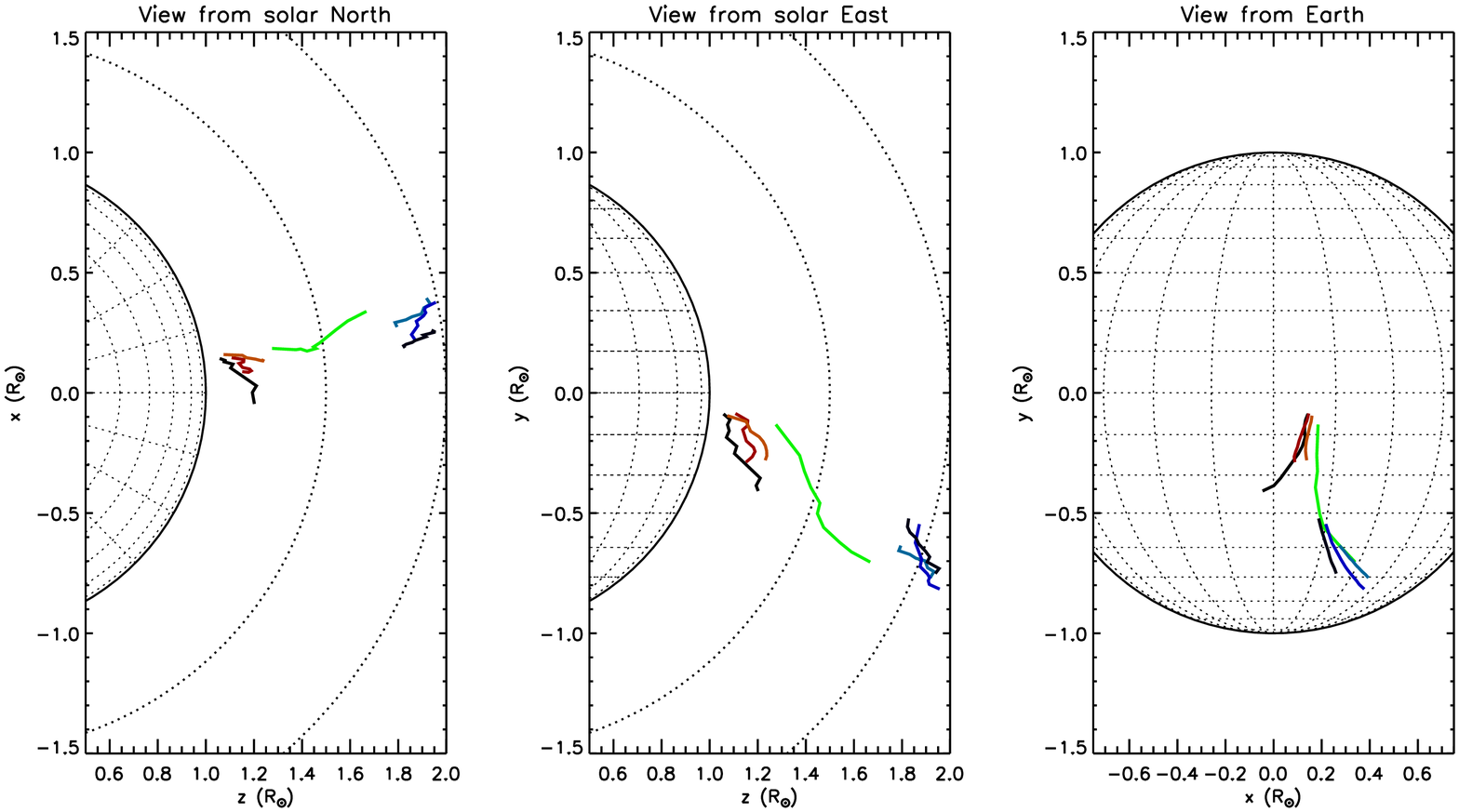}
\caption{3D reconstruction of the erupting filament obtained with tie-pointing technique applied to STEREO/EUVI data. Each colored line represents the filaments identified in different STEREO/EUVI image frames, according to the labels reported in Fig.~\ref{tie_pointing1}. The evolution is plotted as seen from above the ecliptic (left panel), on the ecliptic, from East (middle panel), and from the Earth (right panel).\label{prominence}}
\end{figure}

\begin{figure}[p]
\centering\includegraphics[width=\textwidth]{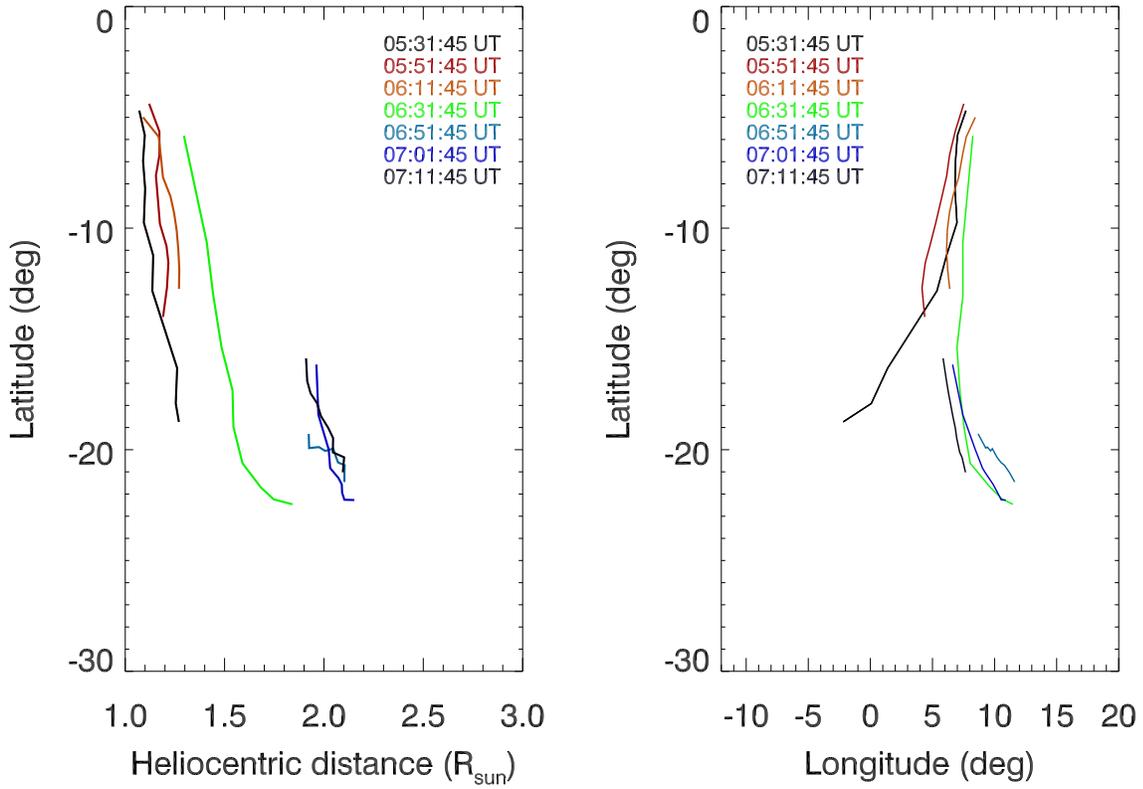}
\caption{Latitude vs. altitude (i.e., side-on; left panel) and altitude vs. longitude (i.e., face-on; right panel) diagrams showing the evolution of the reconstructed filament (see text). Different colors refer to the same filament identified in different STEREO/EUVI image frames according to the times indicated by the labels.\label{tie_pointing1}}
\end{figure}

Figure~\ref{prominence} shows a 3D representation of the reconstructed trajectory of the filament identified in several EUVI frames, as observed from three different prespectives. Detailed latitude vs. altitude (i.e., as seen side-on) and altitude vs. longitude (i.e., as seen face-on) diagrams of the filament trajectory are reported in Figure~\ref{tie_pointing1}.
At the beginning of the eruption (05:31~UT), the filament was laying at a heliocentric distance of $\sim 1.1$~\rsun\ above the solar surface and was mostly aligned with the North-South direction. 
In the following hours, the filament expanded southward and westward, reaching the maximum height of about 2.1\rsun\ at 07:11~UT.
In the last phase of the eruption, between 06:51~UT and 07:01~UT, it underwent a rapid counter-clockwise rotation of about 30$^\circ$ around a vertical axis parallel to the LOS (see right panel of Figure~\ref{tie_pointing1}).
Note that there have been other recent observations of similar rotational dynamics in the low corona \citep[e.g.,][]{lie09,vou11,bem11}. 

\begin{figure}[p]
\centering\includegraphics[width=\textwidth]{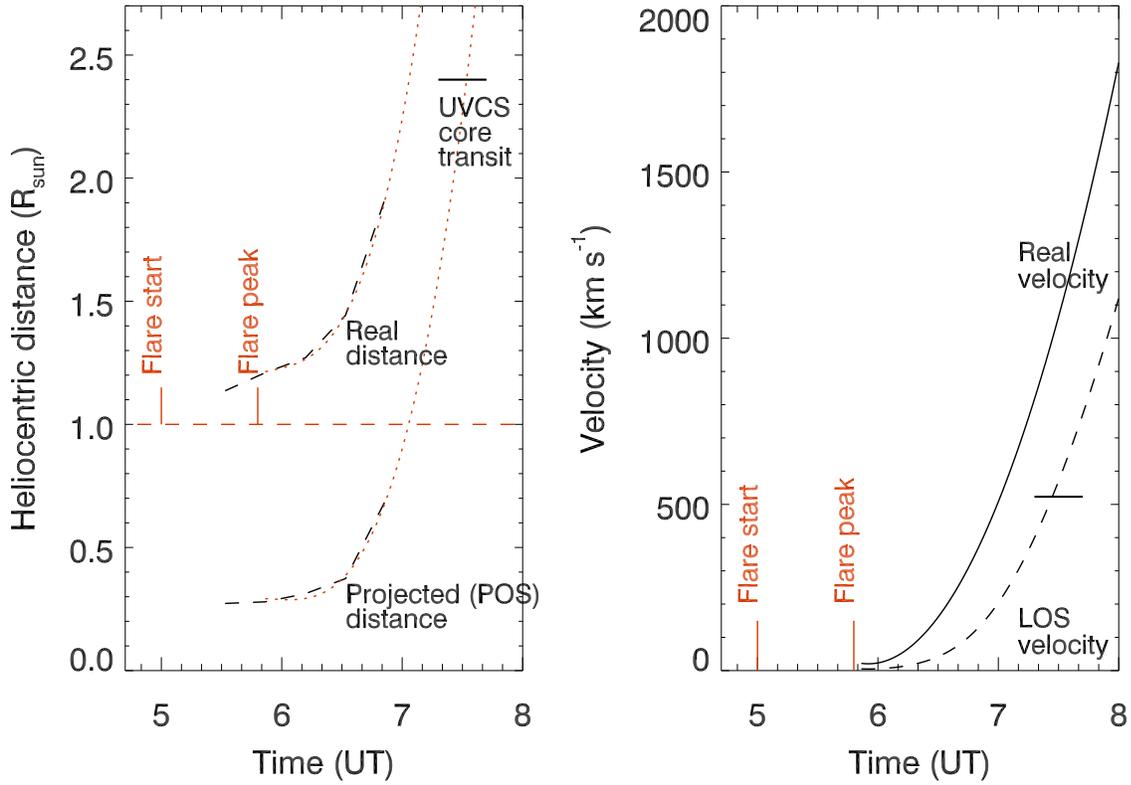}
\caption{Left: temporal evolution of the real and POS projected altitude of the reconstructed filament (dashed lines) and their polynomial extrapolation up to the time of UVCS observations at 2.4~\rsun\ (horizontal solid line). Start and peak time of the solar flare associated with the CME are also shown for comparison purposes. Right: same as left panel, for the real and LOS projected velocity of the filament.\label{tie_pointing2}}
\end{figure}

Figure~\ref{tie_pointing2} shows a polynomial 2nd-order extrapolation of the projected altitude and LOS velocity of the filament at the time of UVCS observations.
The extrapolated quantities match very well the measured ones, in particular the velocity component along the LOS of the reconstructed filament is in very good agreement with the value of $\sim 535$~km~s$^{-1}$ derived from the Doppler shift of \ovi\ lines.
This result suggests that the bright emission structure identified as the CME core in UVCS images (see Figure~\ref{uvcs_int}) actually originates from the prominence plasma embedded in the expanding flux rope.
Note that a convincing identification of the core of the CME with the cool prominence material is relatively rare \citep{mie11}. 

\subsection{3D reconstruction of the CME front with COR1 data}

As already outlined, in the present work we carried out the first ever investigation of the CME properties, by combining the results of the UVCS spectroscopic observations with the geometric information that can be derived from the STEREO images. To perform the 3D reconstruction of the CME expanding front, we applied the polarization ratio technique \citep[see, e.g.,][]{mor04,mie09} to the STEREO/COR1 data.

It is known that the white-light emission in the solar K-corona is originated by Thomson scattering of photospheric light produced by free electrons. Detailed description of the Thomson scattering theory can be found in various papers \citep[e.g.,][]{min30,hul50,bil66}. In particular, it has been demonstrated that the degree of polarization of Thomson-scattered white-light is a function of the scattering angle between the incident light direction and the direction towards the observer \citep{bil66}, that, in turn, can be related through simple geometric considerations to an effective distance of the scattering electron from the POS \citep[see][]{mor04,der05,vou06}. Therefore, the ratio between polarized brightness (pB) and unpolarized brightness (uB) can be used to estimate pixel-by-pixel the location of the scattering coronal plasma along the LOS. 

\begin{figure}[p]
\centering\includegraphics[width=\textwidth]{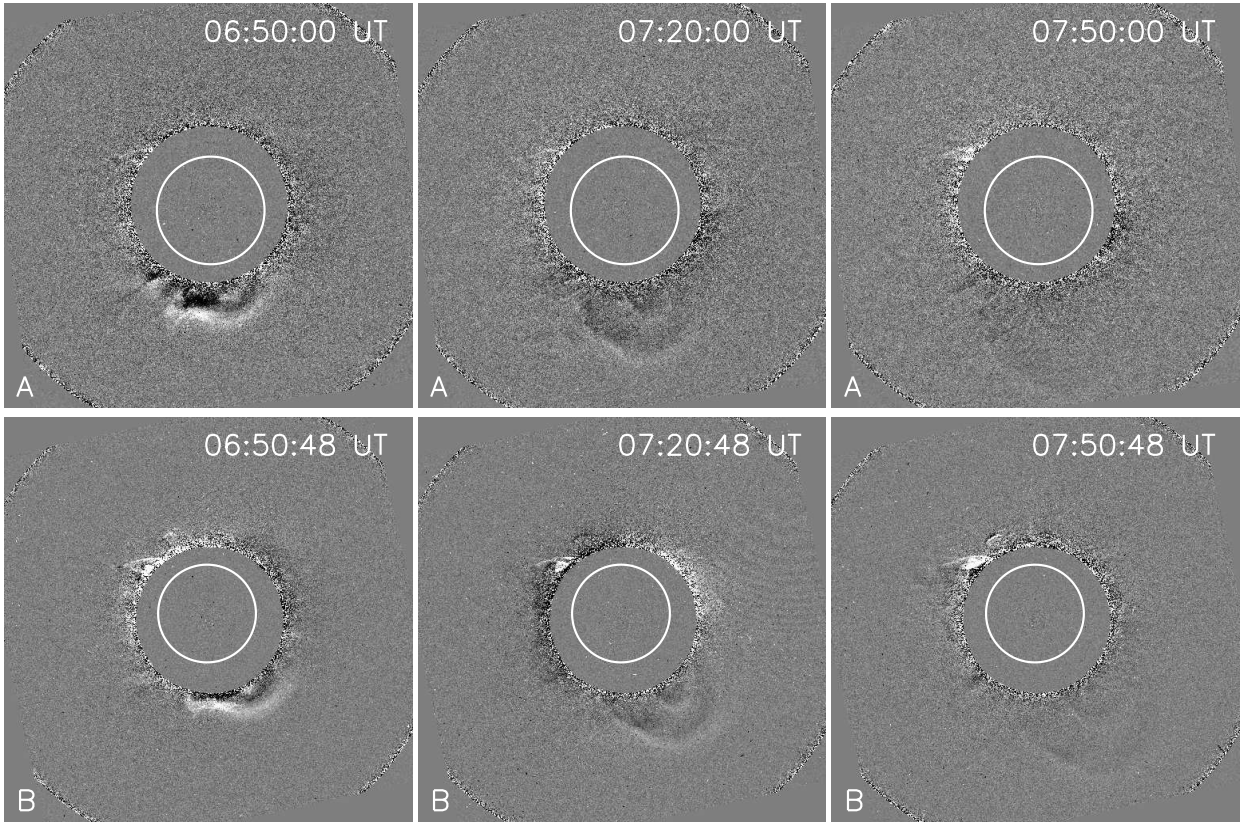}
\caption{Sequence of STEREO/COR1 white-light, base-difference images of the solar corona at three different times during the May 20 CME, as indicated by the labels. Top row is relevant to STEREO~A, bottom row to STEREO~B.\label{cor1}}
\end{figure}

We used total and polarized brightness images obtained by the STEREO/COR1 Ahead and Behind observations of the event. We considered three different times during the expansion of the CME, 06:50~UT, 07:20 UT, and 07:50~UT, which cover the whole transit of the CME through the UVCS field of view.
Figure~\ref{cor1} reports the sequence of base-difference images from STEREO/COR1 at the three considered times, showing the progressive expansion of the CME from the two viewpoints of the Ahead and Behind spacecrafts.
As it can be noted from a comparison with Figure~\ref{lasco}, where the UVCS FOV is superimposed on the images to highlight its position with respect to the CME features, at 07:20~UT the intersection between the CME front and the UVCS field of view was larger, hence, for our discussion, we apply the polarization ratio technique to this time frame.

Recently, \citet{mie11} have shown that the contribution of the H$\alpha$ radiation to the total white-light emission in COR1 coronagraphic images (whose bandpass, at variance with LASCO, also includes the H$\alpha$ $\lambda6563$~\AA\ line) can produce artificial 3D features (called ``H$\alpha$ horns'') in the 3D reconstruction by the polarization ratio technique. 
The H$\alpha$ horns are inconsistent with the real CME structure and they usually appear to be located far from the POS of the coronagraph. 
The reason is that the H$\alpha$ emission coming from the CME core is not Thomson-scattered radiation and results in a polarization decrease of the white-light \citep[see, e.g.,][]{pol76}. 
We needed to exclude this unwanted contribution to determine an accurate 3D CME structure. 
In particular, to prevent the H$\alpha$ contamination, we selected in the COR1 images the area corresponding to the CME front and restricted our analysis to that region.

The ratio pB/uB is calculated pixel-by-pixel and converted into the effective angular distance from the POS along the LOS ($z$ coordinate) by using both STEREO A and B images.
Owing to the forward/backward symmetry of Thomson scattering, the brightness ratio does not indicate whether the scatterer is in front or behind the POS \citep[see][]{mie09}.
However, this ambiguity can be resolved in this case by taking into account that the CME was cataloged as a partial-halo and its source AR was located in the visible hemisphere of the Sun, so that the CME was Earth-directed; therefore, the scattering electron turns out to be located in front of the POS ($z > 0$).

\begin{figure}[p]
\centering\includegraphics[width=0.8\textwidth]{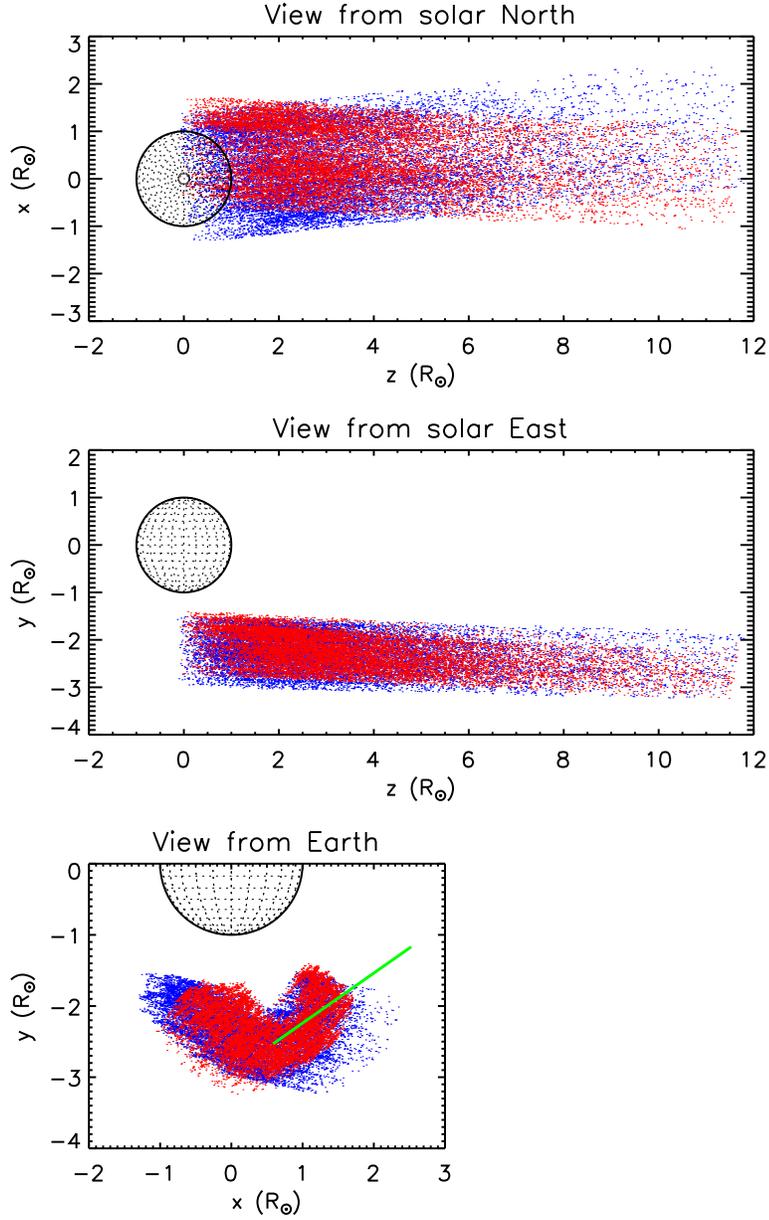}
\caption{3D distribution of the scattering locations identified with the polarization ratio technique applied to STEREO-B (red) and -A (blue) COR1 images at 07:20~UT, as seen from above the ecliptic (top panel), on the ecliptic, from East (middle panel), and from the Earth (bottom panel).\label{pb_ratio}}.
\end{figure}

In Figure~\ref{pb_ratio} we display the results of the 3D reconstruction, as viewed from three different perspectives. 
The blue and red dots represent the 3D distribution of scattering locations for the CME front derived from COR1/STEREO~A and~B pixels, respectively.
The analysis of each COR1 image provided about 18000 points in the 3D coordinate system having the $z$ axis directed towards the Earth, and the $x$ and $y$ axes along the East and North directions, respectively.
It is particularly evident from the top panel in the Figure that each of the two point distributions is mainly concentrated around the direction of the line of sight of the corresponding spacecraft \citep[as already found, e.g., by][]{mie11}.

In order to obtain more detailed information on the 3D structure of the CME front, we applied a statistical analysis to the whole distribution of identified scattering locations.
First, we divided the 3D coordinate system into cubic cells with 0.1~\rsun\ side and measured the number of scattering points included in each cell, $N_{i,j,k}$ (the indexes $i$, $j$, and $k$ denote the position of each cell in the 3D space); then, for each column in the $z$ direction, defined by a particular $(i',j')$ pair, we calculated the average of the resulting $N_{i',j',k}$ distribution, 
\begin{equation}
\langle z \rangle=\frac{\sum_k N_{i',j',k}\cdot z_k}{\sum_k N_{i',j',k}}
\end{equation}
as well as the standard deviation of the distribution, $\Delta z$. Quantities $\langle z \rangle$ and $\Delta z$ have been computed by averaging over all the 3D points obtained by both STEREO-A and -B spacecraft images.

\begin{figure}[p]
\includegraphics[width=\textwidth,trim=0 7cm 0 0,clip=true]{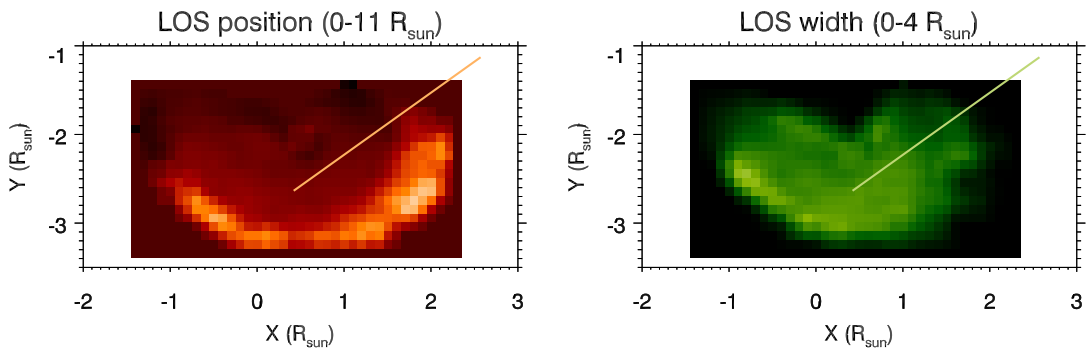}
\caption{Topographical 2D images showing the average distance from the POS ($(x,y)$ plane; left panel) and average depth along the LOS ($z$ axis; right panel) of the CME front, as derived from a statistical analysis applied to the point distribution shown in Figure~\ref{pb_ratio} (see text).\label{front_geom1}}
\end{figure}

The results are reported in Figure~\ref{front_geom1}.
The left panel shows a topographical 2D map of the CME front displaying the average distance of the scattering locations from the $(x,y)$ plane (the quantity $\langle z \rangle$) for each column in the LOS direction. 
The color scale indicates different distances from the POS in solar radii.
The identified points are distributed between 0 and 12~\rsun.
The front turns out to be similar to a ``curtain'' of plasma or a loop arcade system expanding towards the Earth \citep[similar to the results reported by][]{mor04}.
It is worth noting that the greatest distances ($\sim 12$~\rsun) are achieved at the edges of the front.
The right panel reports the corresponding standard deviation of the distribution, that we interpret as the depth along the LOS of the CME front.
The front depth is up to $\sim 4$~\rsun, and the maximum value is achieved in the same regions that appear to be more distant from the POS.

\begin{figure}[p]
\includegraphics[width=\textwidth]{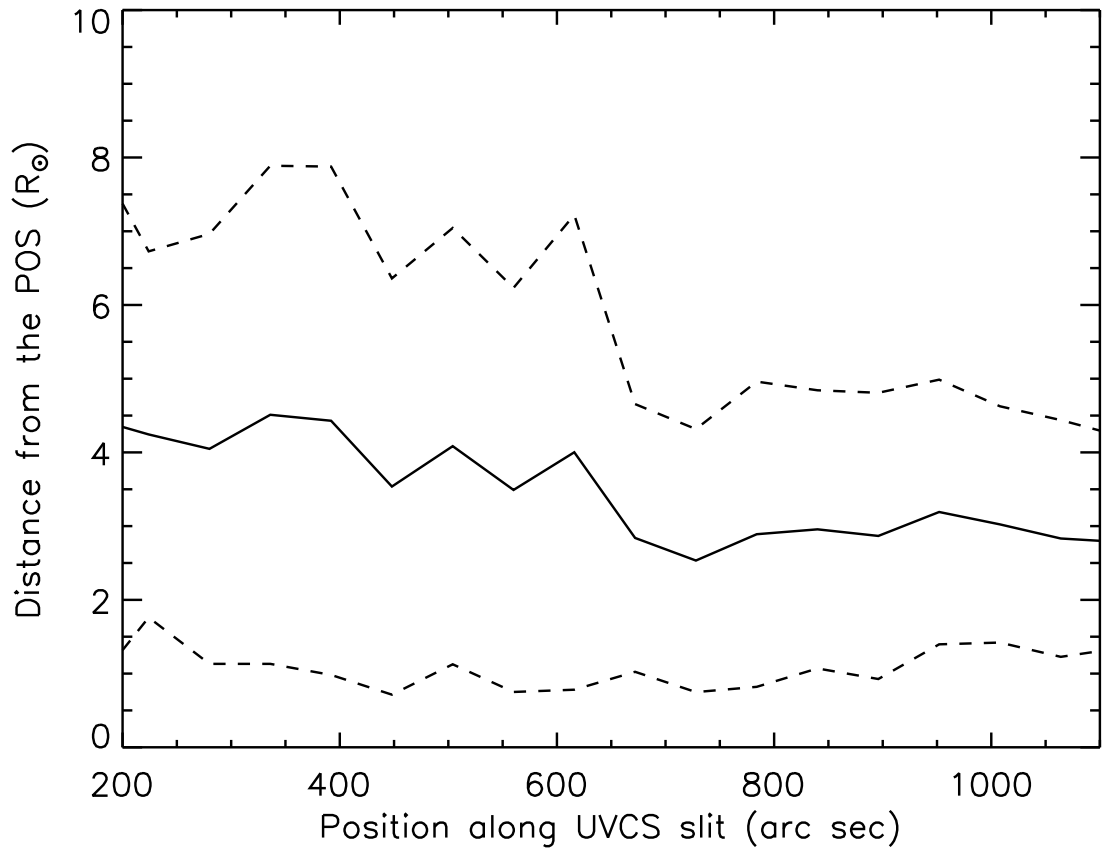}
\caption{Extrapolated average POS distance and LOS depth of the CME front, as derived from the polarization ratio technique applied to STEREO/COR1 data, plotted as function of the position along the UVCS slit, expressed in solar radii.\label{front_geom2}}
\end{figure}

In order to use these geometric parameters in correlation with UVCS data, we extrapolated the mean CME front position and the mean front depth at the location of the UVCS field of view.
Figure~\ref{front_geom2} reports these quantities as functions of the position along the UVCS slit. 
Note that the CME front is located at $\sim 4$~\rsun from the POS at the left edge of the slit, which is located near the center of the front (see Figure~\ref{front_geom2}), and progressively at lower distances from the POS towards the right.
Correspondingly, the LOS depth of the front is maximum at left edge of the slit ($\sim 2$~\rsun), then it decreases.

\subsection{Synthesis of \ovi\ profiles}

Typically, one of the signatures of CME fronts in UV and EUV spectra is the broadening of line profiles that are observed during their transit \citep[see, e.g.,][]{cia05,cia06}.
Both expansion of the emitting volume and increasing kinetic temperature can contribute to broaden the line profiles. The contribution of plasma bulk expansion to the line width at the front can be significant in particular for halo and partial-halo CMEs, due to geometrical effects related to the relative angle between the LOS and the flow velocity. Hence, in order to correctly interpret the measured line widths at the CME fronts, one needs to disentangle bulk expansion and thermal broadening.
\citet{cia06}, for instance, attempted to discriminate between the two components for several observed halo CMEs by assuming a spherical shell model of the expanding front, by estimating the component of the broadening due to bulk expansion using information on the LOS and POS speeds, and by comparing the result with line widths of the \ovi\ spectral profiles measured with UVCS.
Although the authors were able to attribute line broadening to plasma heating in some cases and to the LOS component of the CME expansion in other cases, this approach has some limitations, most related to the assumption of the CME front shape.
In this section we address this issues by combining for the first time the results derived with UVCS spectroscopic data and those derived with STEREO 3D reconstructions.
In particular, the information on the position and depth of the CME front along the LOS derived with the pB-ratio technique applied to STEREO/COR1 images are used for the synthesis of spectral profiles of the \ovi\ $\lambda$1031.91~\AA\ line to be compared with the measured line profiles.

It is well know that, since coronal plasma is optically thin, UV spectra in the solar corona originate from the superimposition along the line of sight of radiation emitted locally by plasma elements characterized by different electron density, electron and kinetic temperatures, and flow velocity.
Therefore, in order to build up a synthetic profile of a particular spectral line, one has to make assumptions on the distribution along the LOS of those plasma quantities.
In our case, we performed a synthesis of the \ovi\ $\lambda$1031.91~\AA\ line considering in particular the lines of sight defined by the spatial bins of the UVCS field of view, limiting the calculation to the portion of the UVCS slit that has been effectively crossed by the front of the CME at 07:20~UT of May 20 (i.e., about 50\% of the slit).
For a given bin (i.e., for a given LOS), we made the simplifying assumption that all plasma elements contributing to the \ovi\ total line profile for that LOS emitted a purely Gaussian spectral profile given by the following expression:
\begin{equation}
P_n(\lambda,z)=I_n(z)\exp\left\{-\left[\frac{\lambda-\lambda_0(1+v_n(z)/c)}{\Delta\lambda_n(z)}\right]^2\right\},
\end{equation}
where the index $n$ denotes the $n$-th bin of the UVCS slit, $\lambda_0=1031.91$~\AA\ is the rest wavelength of the \ovi\ line transition, $c$ is the light speed, $I_n(z)$ and $v_n(z)$ are the assumed functions giving the variation along the LOS of the \ovi\ emitted intensity and flow velocity, and $\Delta\lambda_n(z)$ is the assumed width of the contributing profile, corresponding to an assumption on the kinetic temperature of the O$^{5+}$ ions.

The functions $I_n(z)$ and $v_n(z)$ were set up \emph{ad hoc} and iteratively adjusted until the best agreement between the full width at half maximum (FWHM) of the synthetic profiles and that of the observed profiles was found, within the uncertainties.
The functions giving the best result have the following expressions:
\begin{equation}
I_n(z)=\exp\left[-\left(\frac{z-\langle z \rangle_n}{0.1\cdot\Delta z_n}\right)^2\right],
\end{equation}
\begin{equation}
v_n(z)=\left\{\begin{array}{ll}
\alpha\,m \arctan(z-\langle z \rangle_n)+v_n^{UVCS} & \quad z < \langle z \rangle_n \\
m \arctan(z-\langle z \rangle_n)+v_n^{UVCS}        & \quad z \geq \langle z \rangle_n
\end{array}\right.,
\end{equation}
where $\langle z \rangle_n$ and $\Delta z_n$ are the average position and the width of the CME front derived as explained above for the $n$-th bin along the UVCS slit, $m=-180$~km~s$^{-1}$ is a scale factor, $\alpha =3$ is a parameter that is introduced to take into account asymmetries of the observed spectral profiles, and $v_n^{UVCS}$ is the front velocity measured from the Doppler shift of the \ovi\ $\lambda 1031.91$~\AA\ line using UVCS data.
Note that the peak intensity distribution along the LOS is a Gaussian centered at the position of the CME front and having a $1/e$ full width that is 10\% of the front width; this indicates that the LOS depth of the CME front in UV can be narrower than that derived from white light data, probably because \ovi\ emission is efficient only in the very interior of the front due to plasma density and temperature distributions.
The velocity function is such that plasma elements located ahead (behind) the CME front have flow speeds larger (smaller) than that of the front itself, i.e., the front is not expanding rigidly.

Few considerations must be done about the assumption on the line width of the contributing profiles $\Delta\lambda_n(z)$, i.e., on the kinetic temperature of the emitting plasma elements.
For the sake of simplicity, we assumed that the distribution of line widths is uniform both along the LOS and the UVCS slit, i.e., $\Delta\lambda_n(z)=const.=0.39$~\AA, corresponding to an effective O$^{5+}$ temperature of $\sim 1.2 \times 10^7$~K.
This temperature is quite higher than typical values measured from UV data in the front of CMEs \citep[$< 10$~MK; see, e.g.,][]{bem07}, but it is $\sim 70$\% lower than the temperature measured from data analyzed here in the undisturbed pre-CME corona at the same altitude ($\sim 1.7 \times 10^7$~K, as reported in \paperone).
It is worth noting that the assumption of uniform temperature along the LOS could not be representative of the real solar corona due to the possible superimposition along the LOS of different structures characterized by different temperatures; in our case, however, the bulk of the \ovi\ emission comes from a relatively narrow region ($\sim 0.4$~\rsun wide) centered around the CME front, as explained above, so this assumption can be considered fairly valid.

The total synthetic profile for a given bin of the UVCS field of view is obtained by integration along the LOS of the elemental contributing profile:
\begin{equation}
P_n(\lambda)=\int P_n(z,\lambda)\,\mathrm{d}z;
\end{equation}
this integral is turned into a sum considering that the $z$ coordinate is a discretized variable.
The FWHM of the total profile is then compared as function of the position along the UVCS slit, with the measured FWHM of the observed \ovi\ profiles.
In order to enhance the statistics and increase the signal-to-noise ratio, we accumulated the observed profiles over 8 exposures (corresponding to a 16~mins average) around 07:20~UT.
Line profiles were integrated and the measured FWHMs were corrected for instrumental broadening as usual \citep[see, e.g.,][]{koh99}.

\begin{figure}[p]
\includegraphics[width=\textwidth]{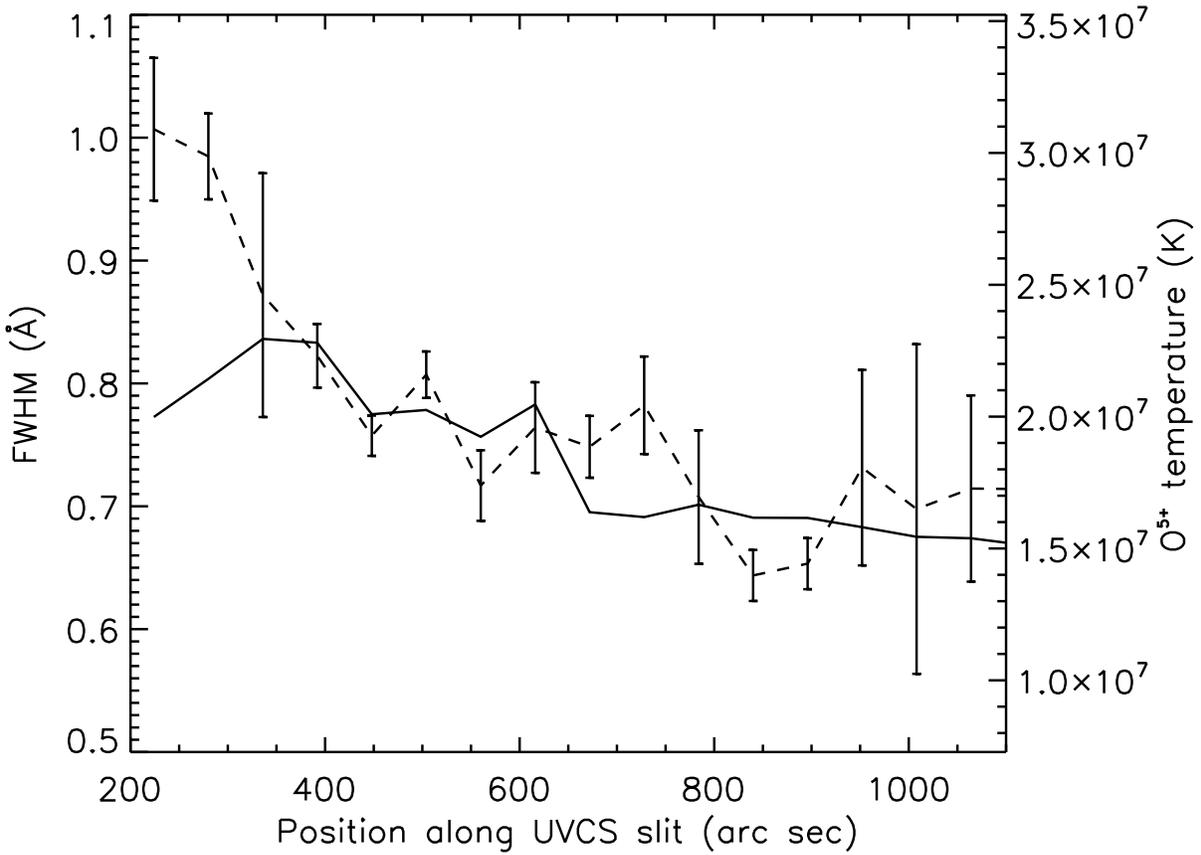}
\caption{FWHM of the observed (dashed line) and synthesized (solid line) profiles of the \ovi\ $\lambda$1031.91~\AA\ line, as functions of the position along the UVCS slit.\label{fwhm}}
\end{figure}

The resulting observed and synthetic FWHMs are shown in the plot of Figure~\ref{fwhm} as function of the heliocentric distance along the UVCS slit.
Note that points at higher altitude are located at the front center, while points at lower altitude are located in the peripheral part of the front (see, for comparison, Figure~\ref{front_geom1}).
The uncertainties in the observed FWHMs were computed with the standard error propagation formulae and are plotted as error bars in the Figure.
The observed line widths are well reproduced (within the uncertainties) by the synthetic ones only in the peripheral part of the CME front.
At the front center there is no agreement, and the observed FWHMs are $\sim 25$\% higher than the synthetic ones.
Any other intensity and velocity distribution along the LOS able to reproduce the FWHMs observed at the front center was unable to reproduce at the same time the FWHMs measured at other locations along the slit.

When the FWHM is converted into effective temperature of the O$^{5+}$ ions, it turns out that the observed line widths imply a temperature that is $\sim 1.6$ times higher than the temperature corresponding to the synthetic line widths.
The inability of reproducing the observed line widths at the front center through the synthesis of spectral profiles that takes into account both thermal and non-thermal components of the line widths suggests that, at least at the front center, physical mechanisms other than bulk expansion are responsible for line broadening and provide the additional heating that is necessary to increase the plasma temperature up to the observed values.
One of the possible mechanisms is represented by plasma compression that most likely occurs in the central part of the CME front due to the relatively rapid expansion of the eruption in the pre-CME medium.
It is worth noting that observed line widths at the peripheral part of the CME front were reproduced by assuming a 70\% lower kinetic temperature with respect to the pre-CME corona. This can indicate that the CME front is made of plasma dragged by the expansion of the flux rope and coming from lower layers of the solar corona, characterized by lower temperatures. This association between the CME front plasma and the ambient coronal plasma dragged by the outward expansion of the CME core is a common characteristic of several CME models (e.g., the flux-rope models or the breakout models; see Forbes, 2006), and our results can provide clear evidence for that.

\section{Conclusions}

In this work we carried out the first ever analysis of a partial-halo CME by combining the results from the UVCS spectroscopic observations with the 3D geometric information that can be derived from the STEREO data. We studied the first CME observed simultaneously by STEREO and SOHO/UVCS: this event occurred on 2007, May 20 and was visible from about 04:30~UT to 09:00~UT. 
We performed 3D reconstructions of both the erupting prominence associated with the CME and the CME front using EUVI and COR1 data. 
Then, we exploited the geometric information derived from the reconstructed CME structure to synthesize spectral profiles of the \ovi\ $\lambda$1031.91~\AA\ line and compared them with spectral profiles observed by UVCS during the transit of the CME.

The main results of this work can be summarized as follows:
\begin{enumerate}
\item We applied the tie-pointing technique to the STEREO/EUVI~304~\AA\ images of the CME source active region in order to reconstruct the 3D kinematics of the erupting filament. During the eruption, between 06:51~UT and 07:01~UT of May 20, the filament, which was initially oriented in the north-south direction, underwent a rapid counter-clockwise rotation of about 30$^\circ$ around a vertical axis parallel to the LOS.

\item A comparison of a polynomial extrapolation of the projected altitude and LOS velocity of the filament at the time of UVCS observations with the altitude of the UVCS field of view and the LOS velocity component derived for the CME core suggests that the bright emission structure identified as the CME core in white-light images and UV data actually originates from the prominence plasma embedded in the expanding flux rope.

\item We applied the polarization ratio technique to the STEREO/COR1-A and -B data at 07:20~UT of May 20 in order to perform the 3D reconstruction of the CME expanding front. The scattering locations identified with this method are distributed between 0 and 12~\rsun from the POS. The CME front turns out to be similar to a ``curtain'' of plasma or a loop arcade system expanding towards the Earth with the greatest distances achieved at the edges of the front. The LOS depth of the front is up to $\sim 4$~\rsun; the front appears wider at the center, where it is also more distant from the POS.

\item We used the geometric information derived by the polarization-ratio analysis to synthesize spectral profiles of the \ovi\ $\lambda$1031.91~\AA\ line expected at different positions along the UVCS field of view and taking into account bulk expansion of the emitting volume by assuming an intensity and velocity distribution of the plasma elements along the LOS and a constant kinetic temperature of the O$^{5+}$ ions.

\item The synthetic line widths well reproduce (within the uncertainties) the observed ones only in the peripheral part of the CME front, while at the front center, where the distance from the POS is greater, they turn out to be $\sim 25$\% lower than the others. By converting the synthetic line widths into effective temperatures, the front center is characterized by a temperature $\sim 1.6$~times higher than that of the peripheral regions. This indicates the presence of a significant heating mechanism responsible for line broadening other than bulk expansion and LOS integration. 
\end{enumerate}

The occurrence of plasma heating is, together with the detection of type-II radio bursts, a good indication of CME-driven shock propagation. 
In particular, plasma heating detected here is an increase by a factor $\sim 1.6$ in the O$^{5+}$ ion kinetic temperatures at the center of the CME front. 
On the other hand, this event was not associated with any type-II radio burst, and this indicates that no shock was excited by the expansion of the CME front. 
This is also in agreement with the quite low front velocity we derived: when the front was at the unprojected heliocentric distance of ($2.4$~\rsun)$/\sin(30^\circ)=4.8$~\rsun, its velocity was 578~km~s$^{-1}$, hence smaller than local Alfv\'en velocities ($\sim 700$~km~s$^{-1}$) expected at these altitudes \citep[see, e.g.][]{mann03}. 
Hence, we may argue that plasma heating inferred here is simply due to plasma compression: for instance, under the assumption of adiabatic compression of a perfect gas, a temperature increase by a factor $K_T = 1.6$ corresponds to a density increase by a factor $K_n = K_T^{1/(\gamma-1)}\simeq 2.0$, which is quite a reasonable value. 
Interestingly, the lack of significant plasma heating we detect at the peripheral region of the front indicates that stronger plasma compression is occurring only at the front center. 
This behaviour is also similar to what recently found by \citet[][]{bemporad14} for the adiabatic heating occurring across a CME-driven shock.

Overall, this work demonstrates the potential of combining 3D kinematical information from stereoscopic and polarized data with spectroscopic observations providing a new capability for a better determination of the CME 3D structure, unprojected expansion velocity and associated plasma heating. Unfortunately, the lack of UV spectrometers of the extended corona (like SOHO/UVCS) in the payload of the next accepted solar missions will prevent the possibility to apply the analysis described here to future data.
Alternative possibilities could be offered by future coronagraphs whose fields of view will extend well below 2~\rsun, thus overlapping with spectroscopic observations provided by future solar missions (like Solar Orbiter/SPICE).

\acknowledgments
\begin{acknowledgements}
This work has received funding from the European Commission's Seventh Framework Programme (FP7/2007-2013) under the grant agreement SWIFF (project no. 263340, {\tt www.swiff.eu}) and from the Agenzia Spaziale Italiana through the contract ASI/INAF no. I/013/12/0.
\end{acknowledgements}

\newpage

\newpage

\begin{thebibliography}{}
\bibitem[Antonucci(2006)]{ant06} Antonucci, E. 2006, Space Sci. Rev., 124, 35

\bibitem[Bemporad(2007)]{bem07} Bemporad, A., Raymond, J. C., Poletto, G., \& Romoli, M. 2007, ApJ, 655, 576

\bibitem[Bemporad et al.(2011)]{bem11} Bemporad, A., Mierla, M., \& Tripathi, D. 2011, A\&A, 531A, 147B

\bibitem[Bemporad, Susino \& Lapenta(2014)]{bemporad14} Bemporad, A., Susino, R., \& Lapenta, G. 2014, ApJ, 784, 102

\bibitem[Billings(1966)]{bil66} Billings, D. E., 1966, \emph{A Guide to the Solar Corona}, Academic Press, London

\bibitem[Bothmer(2003)]{bot03} Bothmer, V. in: Wilson, A. (Ed.), International Solar Cycle Studies (ISCS) Symposium. ESA SP-535. Noordwijk: ESA Publications Division, ISBN 92-9092-845-X, 419, 2003

\bibitem[Bradshaw \& Raymond(2013)]{bra13} Bradshaw, S.J. \& Raymond, J.C. 2013, Space Sci. Rev.

\bibitem[Brueckner et al.(1995)]{bru95} Brueckner, G. E., Howard, R. A., Koomen, M. J., et al. 1995, Sol. Phys., 162, 357

\bibitem[Burkepile et al.(2004)]{bur04} Burkepile, J. T., Hundhausen, A. J., Stanger, A. L., et al. 2004, J. Geophys. Res., 109, 3103

\bibitem[Ciaravella et al.(2005)]{cia05} Ciaravella, A., Raymond, J. C., Kahler, S., et al. 2005, ApJ, 621, 1121

\bibitem[Ciaravella et al.(2006)]{cia06} Ciaravella, A., Raymond, J. C., \& Kahler, S. W. 2006, ApJ, 652, 774

\bibitem[Cremades \& Bothmer(2004)]{cre04} Cremades, H., \& Bothmer, V. 2004, A\&A, 422, 307

\bibitem[Dere et al.(2005)]{der05} Dere, K. P., Wang, D., \& Howard, R. 2005, ApJ, 620, 119

\bibitem[Giordano et al.(2013)]{gio13} Giordano, S., Ciaravella, A., Raymond, J. C., et al. 2013, J. Geophys. Res., 118, 967

\bibitem[Gopalswamy et al.(2009)]{gop09} Gopalswamy, N., et al. 2009, Earth Moon Planets, 104, 295

\bibitem[Gosling(1993)]{gos93} Gosling, J. T. 1993, J. Geophys. Res., 98, 937

\bibitem[Hu \& Sonnerup(2002)]{hu02} Hu, Q., \& Sonnerup, B. U. O. 2002, J. Geophys. Res., 107

\bibitem[van de Hulst(1950)]{hul50} van de Hulst, H. C. 1950, Bull. Astron. Inst. Neth., 11, 135

\bibitem[Huttunen et al.(2002)]{hut02} Huttunen, K. E. J., Koskinen, H. E. J., \& Schwenn, R. 2002, J. Geophys. Res., 107, 1121

\bibitem[Inhester(2006)]{inh06} Inhester, B., 2006, arXiv:astro-ph/0612649

\bibitem[Kaiser et al.(2008)]{kai08} Kaiser, M. L., Kucera, T. A., Davila, J. M., et al. 2008, Space Sci. Rev., 136, 5

\bibitem[Kilpua et al.(2009)]{kil09} Kilpua, E. K. J., Liewer, P. C., Farrugia, C., et al. 2009, Sol. Phys., 254, 325

\bibitem[Kohl \& Withbroe(1982)]{koh82} Kohl, J. L., \& Withbroe, G. L. 1982, ApJ, 256, 263

\bibitem[Kohl et al.(1995)]{koh95} Kohl, J. L., et al. 1995, Sol. Phys., 162, 313

\bibitem[Kohl et al.(1999)]{koh99} Kohl, J. L., Esser, R., Cranmer, S. R., et al. 1999, ApJ, 510, L59

\bibitem[Kohl et al.(2006)]{koh06} Kohl, J. L., Noci, G., Cranmer, S. R., \& Raymond, J. C. 2006, Astron. Astrophys. Rev., 13, 31

\bibitem[Lanzerotti et al.(2001)]{lan01} Lanzerotti, L. in: \emph{Space Weather Effects on Technologies}, ed. by P. Song, H.J. Singer, G. L. Siscoe, Space Weather, AGU Monograph 125, 11, 2001

\bibitem[Lee et al.(2006)]{lee06} Lee, J.-Y., Raymond, J. C., Ko, Y.-K., \& Kim, K.-S. 2006, ApJ, 651

\bibitem[Liewer et al.(2009)]{lie09} Liewer, P. C., De Jong, E. M., Hall, J. R., et al. 2009, Sol. Phys. 256, 57

\bibitem[Mann et al.(2003)]{mann03} G. Mann, G., Klassen, A., Aurass, H., \& Classen, H.-T. 2003, A\&A, 400, 329

\bibitem[Mierla et al.(2008)]{mie08} Mierla, M., Davila, J., Thompson, W., et al. 2008, Sol. Phys., 252, 385

\bibitem[Mierla et al.(2009)]{mie09} Mierla, M., Inhester, B., Marqu\'e, C., et al. 2009, Sol. Phys., 259, 123

\bibitem[Mierla et al.(2010)]{mie10} Mierla, M., Inhester, B., Antunes, A., et al. 2010, Ann. Geophys., 28, 203

\bibitem[Mierla et al.(2011)]{mie11} Mierla, M., Chifu, I., Inhester, B., et al. 2011, A\&A, 530, L1

\bibitem[Minnaert(1930)]{min30} Minnaert, M. 1930, Z. Astrophys. 1, 209

\bibitem[Moran \& Davila(2004)]{mor04} Moran, T. G. \& Davila, J. M. 2004, Science, 305, 66

\bibitem[Morgan(2012)]{mor12} Morgan, H., Byrne, J. P., \& Habbal, S. R. 2012, ApJ, 752, 144

\bibitem[Noci, Kohl, \& Withbroe(1987)]{noc87} Noci, G., Kohl, J. L., \& Withbroe, G. L. 1987, ApJ, 315, 706

\bibitem[Poland \& Munro(1976)]{pol76} Poland, A. I., \& Munro, R. H. 1976, ApJ, 209, 927

\bibitem[Schwenn et al.(2006)]{sch06} Schwenn, R., Raymond, J. C., Alexander, D., et al. 2006, Space Sci. Rev., 123, 127

\bibitem[Susino et al.(2013)]{sus13} Susino, R., Bemporad, A., Dolei, S., \&  Vourlidas, A. 2013, Adv. Space. Res., 52, 967

\bibitem[Thernisien, Howard, \& Vourlidas(2006)]{the06} Thernisien, A. F. R., Howard, R. A., \& Vourlidas, A. 2006, ApJ, 652, 763

\bibitem[Thernisien, Vourlidas, \& Howard(2009)]{the09} Thernisien, A. F. R., Vourlidas, A., \& Howard, R. A. 2009, Sol. Phys, 256, 111 

\bibitem[Tsurutani et al.(1988)]{tsu88} Tsurutani, B. T., Gonzalez, W. D., Tang, F., et al. 1988, J. Geophys. Res., 93, 8519

\bibitem[Vourlidas \& Howard(2006)]{vou06} Vourlidas, A., \& Howard, R. A. 2006, ApJ, 642, 1216

\bibitem[Vourlidas et al.(2011)]{vou11} Vourlidas, A., Colaninno, R., Nieves-Chinchilla, T., \& Stenborg, G. 2011, ApJ, 733, L23

\bibitem[Webb(1988)]{web88} Webb, D. F. 1988, J. Geophys. Res., 93, 1749

\bibitem[Webb \& Howard(2012)]{web12} Webb, D. F. \& Howard, T. A. 2012, Living Rev. Sol. Phys., 9, 3

\bibitem[Withbroe et al.(1982)]{wit82} Withbroe, G. L., Kohl, J. L., Weiser, H, \& Munro, R. H. 1982, Space Sci. Rev., 33, 17

\bibitem[Yurchyshyn et al.(2001)]{yur01} Yurchyshyn, V. B., Wang, H., Goode, P. R., Deng, Y. 2001, ApJ, 563, 381

\bibitem[Zhang et al.(2007)]{zha07} Zhang, J., Richardson, I. G., Webb, D. F., et al. 2007, J. Geophys. Res., 112
\end{thebibliography}
\end{document}